\documentclass[referee]{raa}           
\usepackage{graphicx,times}
\usepackage{natbib}
\usepackage{amssymb,amsmath}
\bibpunct{(}{)}{;}{a}{}{,}

\usepackage{color}
\usepackage{array}
\usepackage{color}
\usepackage{pifont}
\usepackage{csquotes}
\bibpunct{(}{)}{;}{a}{,}{,}

\newcommand{\degree}{\ensuremath{^\circ}}

\usepackage[a4paper, total={6in, 10in}]{geometry}

\usepackage[a4paper=true,pagebackref=true] {hyperref}
\hypersetup{colorlinks = true, linkcolor = green, anchorcolor = red, citecolor = blue, filecolor = red, pagecolor = red, urlcolor = red}

\begin{document}

   \title{On the photo-evaporation, dust polarization, and kinematics of two nebulae in Sh2-236}

 \volnopage{ {\bf 20XX} Vol.\ {\bf X} No. {\bf XX}, 000--000}
   \setcounter{page}{1}

   \author{Archana Soam\inst{1,2}
   }

   \institute{SOFIA Science Center, Universities Space Research Association, NASA Ames Research Center, M.S. N232-12, Moffett Field, CA 94035, USA; {\it asoam@usra.edu}\\
        \and
             Korea Astronomy and Space Science Institute, 776 Daedeokdae-ro, Yuseong-gu, Daejeon, Republic of Korea.\\
\vs \no
   {\small Received 2020 July 23; accepted 2020 Oct 08}
}

\abstract{In the work presented here, the impact of magnetic field on the dynamical evolution of cometary globules Sim 129 and Sim 130 in Sh2-236 H\,II region and the ionized gas streaming out of their surfaces is investigated. The magnetic field morphology in the region associated with these globules is inferred using optical polarization measurements with the Sampurnanand Telescope at ARIES. The nebular emission is probed through radio continuum mapping at 1.4~GHz with the archival National Radio Astronomy Observatory (NRAO) very large array (VLA) Sky Survey (NVSS) data. The correlation of these measurements suggest that the photoevaporated gas from the surfaces of Sim 129 and Sim 130 is accumulated in clouds and starts streaming along the magnetic field lines. The $\rm ^{12}CO$ (J=1-0) molecular line observations are performed towards NGC\,1893 from 14-m single dish radio telescope in Taeduk Radio Astronomy Observatory (TRAO). The velocity dispersion in $\rm ^{12}CO$ (J=1-0) molecular line and the dispersion in polarization angles are used in Davis-Chandrasekhar-Fermi formulation to estimate the magnetic field strength towards two sim nebulae. The average value of field strength is found to be $\sim$60$\mu$G with uncertainty of 0.5 times the estimated value.
}

   \authorrunning{Soam A.}            
   \titlerunning{On the photo-evaporation, dust polarization, and kinematics of two nebulae in Sh2-236}  
   \maketitle

%
\section{Introduction}
\label{sec:intro}
In the presence of magnetic field, the dynamical evolution of the globules and the ionized gas streaming out of their surfaces get modified significantly depending on the strength of magnetic field and its orientation with respect to the incoming ionizing photons \citep{2009MNRAS.398..157H}. This is understandable because, in the ideal magnetohydrodynamics (MHD) limit, matter is well coupled to the magnetic field lines and when the field is sufficiently strong, the magnetic pressure and tension strongly resist any movement of gas in the direction perpendicular to the field orientations. However, no such resistance is offered for the gas motion along the field lines because ionized species can easily move along fields lines.  \citet{2009MNRAS.398..157H} presented the first 3-D MHD simulations towards magnetized globules and found that photoevaporating globules will evolve into more flattened sheet like structures compared to the non-magnetic cases when the cloud initially has a strong B-field (i.e. 100 times the thermal pressure) perpendicular to the UV radiation direction. The photoevaporated gas accumulates between the ionizing source and the cloud making a recombination region which further prevents the supply of ionizing flux reaching the globule from the ionizing source. Studies have attempted to explain this scenario through MHD simulations \citep{2011MNRAS.412.2079M}. These simulations can be tested by tracing the magnetic field orientation with respect to the ionizing radiation in the nebulae sitting on the periphery of HII regions \citep{2018MNRAS.476.4782S}. Assuming the gas to be totally ionized, the total pressure of the ionized layer on the surface of the globule with respect to the internal neutral gas can be estimated. 

These parameters hold the potential to provide useful insight into the dynamical states of the globules in the HII regions and enable us to understand the radiation driven implosion (RDI) process. RDI is a process in which radiation from the massive stars drives an implosion into the globules sitting in the vicinity. This process drives the convergent shock into the globule and causes it to implode due to compression. Using molecular line measurements in periphery of H\,II regions, we can investigate the kinematics of the gas in those region. Polarization and radio spectroscopic observations together provide opportunity to magnetic field structure and associated gas kinematics.

In this work, I studied two nebulae Sim 129 and Sim 130, in the vicinity of a young open cluster NGC\,1893 located at the center of HII region Sh2-236 (a.k.a. IC\,410 ). This region is populated by several massive O and B type stars \citep{2007MNRAS.379.1237M}. A detailed list of the high mass stars and their type is presented by \citep{2007MNRAS.379.1237M}.  Among the listed five O type stars namely HD 242926 (BD\,$+$33 1024, LSV\,$+33^\circ$16), NGC\,1893 HOAG4 (LS $+33^\circ$15), HD\,242935 (BD\,$+$33 1026), BD $+$33\,1025 (LSV $+33^\circ$17, NGC 1893 HOAG5) and HD\,242908 (NGC\,1893 HOAG5, BD\,$+$33 1023), the central O7.5V type star \citep{2007A&A...471..485N} HD\,242935 is found to be the potential candidate responsible for ionizing the surrounding medium.

Here I present the magnetic field geometry of NGC 1893 using optical R-band polarimetry. NIR (H-band) polarimetric observation towards this object have also been carried out using SIRPOL (in prep. by Eswaraiah et al). This region is with smaller angular size ($\sim$10 arcmin) at a distance of 3.2\,kpc which makes it easier to cover in few pointings only. This enables us to cover the whole region by observing a few fields. At the above quoted distance of this cluster, the extent of the two nebulae Sim 129 and Sim 130 is of the order of $\sim 4.5$ pc (with a diameter of head $\sim1.0$ pc) whereas the distance between the nebulae and the ionizing source is $\sim 6$ pc. The morphology of the nebular emission and its orientation with respect to the magnetic field geometry is probed using the archival NVSS data in 1.4~GHz. The estimation of photo-ionizing flux impinging on the two nebulae and to compare it with the expected flux from the O-type stars located towards the region, has also been performed. This paper is structured as follows. Section 2 elaborates the observations and data reduction procedures and the results are presented in section 3. The analysis of the results and the summary of this study are discussed in sections 4 and 5, respectively.

\section{Observations \& Data Reduction}\label{observe}

\subsection{Optical polarization observations using AIMPOL}
Polarimetric observations were carried out with the 104-cm Sampurnanand telescope of Aryabhatta Research Institute of Observational Sciences (ARIES), Nainital, India using Aries IMaging POLarimeter (AIMPOL). I used R-band (standard Jhonson $R_{c}$ filter having  $\lambda_{R_{eff}}$=0.630$\mu$m) in polarimetric observations of two field towards NGC 1893. The polarimeter has a half-wave plate (HWP) modulator and a Wallaston prism. The data reduction procedure with details of image size, CCD characteristics, pixel size and calculation of Stokes parameters at four different angles of HWP are explained in my earlier work \citep{2013MNRAS.432.1502S, 2015A&A...573A..34S}. The instrumental polarization has been removed from the observed target stars by observing the unpolarized standards given in \citet{1992AJ....104.1563S}.

\subsection{Molecular line observations from TRAO}

On-The-Fly (OTF) mapping observations covering a region of $28^{'}\times28^{'}$ towards NGC 1893 were carried out on March 04 and 05, 2016 for the molecular transitions CO (J=1-0) and C$^{18}O (J=1-0)$ (with $V_{LSR} =-7.0~kms^{-1}$) at Taeduk Radio Astronomical Observatory (TRAO). This is a 13.7 m single dish mm wave telescope with SEcond QUabbin Observatory Imaging Array (SEQUOIA) array at Korea Astronomy \& Space Science Institute, South Korea. SEQUOIA has an array of 4$\times$4 pixels and operates in 85$-$115 GHz frequency range. The system temperature is generally found to be 250 K (85$-$110 GHz) to 500 K (115 GHz; $^{12}$CO). SEQUOIA provides the facilities of simultaneous observations in two different tracers within a frequency band of 15 GHz. The beam size (HPBW) and main-beam efficiency in CO is found to be $44^{''}$ and 51$\pm$2\%  \citep{2018ApJS..234...28L}. I observed the target for $\sim$ 180 minutes to achieve a rms of 0.3 K in ${T_{A}}^{\ast}$ scale in CO. Observations of an emission free region within $2^{\circ}$ field of the target coordinates were used to subtract the sky signal. The signal-to-noise ratio is estimated to be $\sim$16 at the brightest position of ${T_{A}}^{\ast}\sim 4.8$ K. The achieved velocity resolution is $\sim 0.25$ km$~s^{-1}$. SiO maser was used for pointing and focusing of the telescope. The pointing of the telescope was as good as $\sim~5-7^{''}$. Data reduction was performed using {\scriptsize CLASS} from {\scriptsize GILDAS}\footnote{Grenoble Image and Line Data Analysis Software; https://www.iram.fr/IRAMFR/GILDAS/}.

\section{Results}

\subsection{Optical polarimetric results}

\begin{figure*}
\begin{center}$
\begin{array}{cc}
\resizebox{7.0cm}{6.5cm}{\includegraphics{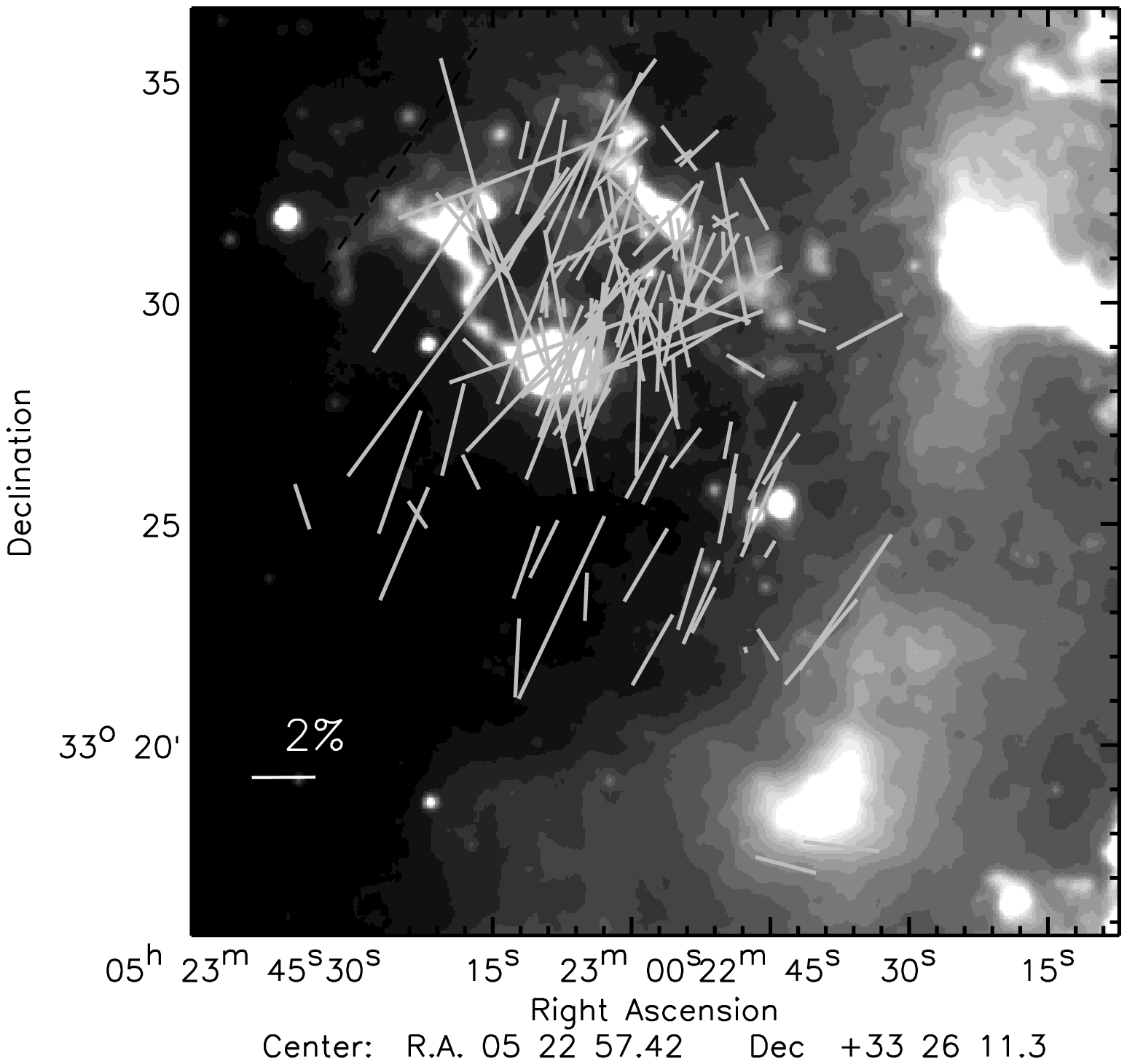}} &
\resizebox{7.5cm}{6.0cm}{\includegraphics{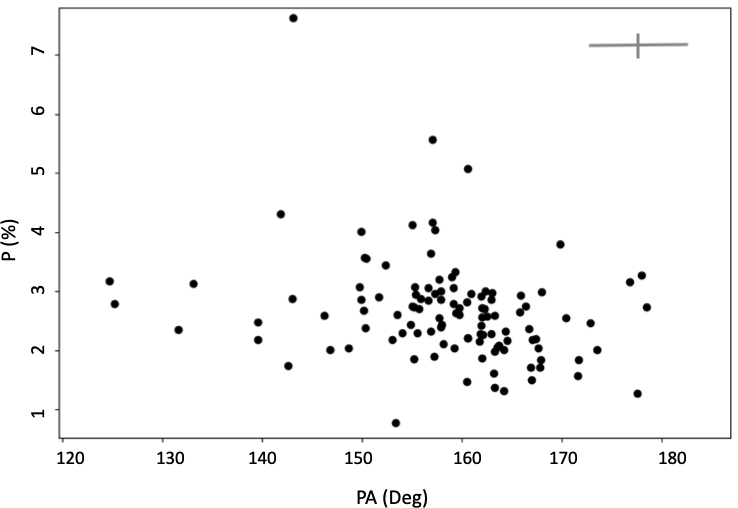}}
\end{array}$
\end{center}
\caption{{\bf Left panel:} The polarization vectors plotted after correcting for the interstellar polarization. {\bf Right panel:}. The P vs $\theta_{P}$ distribution of measurements. A symbol with typical errors in P and $\theta_{P}$ values is also shown.}\label{Fig:corrected}
\end{figure*}

\begin{figure}
\begin{center}$
\begin{array}{cc}
\resizebox{7.8cm}{6.5cm}{\includegraphics{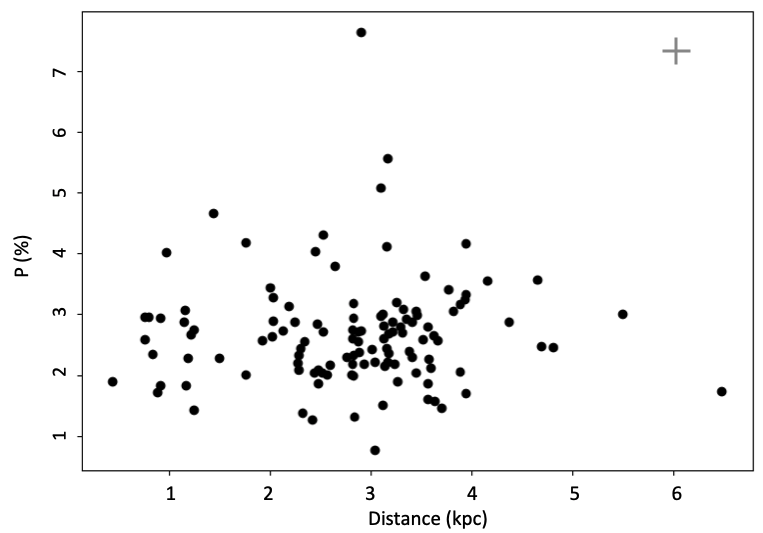}} &
\resizebox{7.8cm}{6.5cm}{\includegraphics{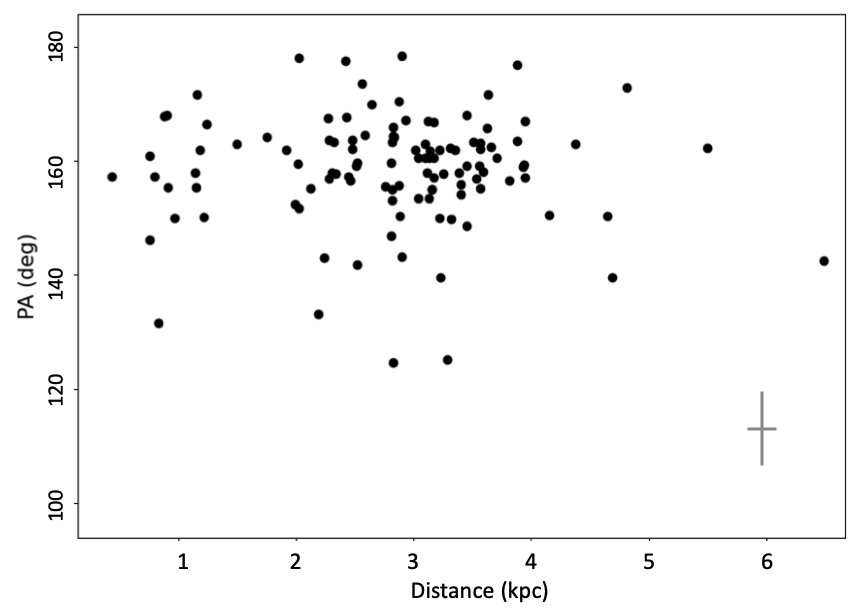}}
\end{array}$
\end{center}
\caption{Left and right panels show variations of P and PA with distances (GAIA DR-2; \citealt{Bailer-Jones2018}) of the stars.}\label{Fig:P_PA_dist}
\end{figure}

The results of optical polarization of 123 sources observed (combining our measurements and previous polarimetric results given by \citet{2011MNRAS.411.1418E} towards NGC 1893 are discussed. \citet{2011MNRAS.411.1418E} observed 44 stars in V-band. Our polarization angles are consistent with those of their results. The nebula Sim 130 was not covered in the previous polarization observations by \citet{2011MNRAS.411.1418E}. In this work, I covered the two nebulae by imaging deeper into the cloud and increased the sample to get a relatively high spatial resolution magnetic field geometry in the region between HD 242935 and the two nebulae. The range of degree of polarization (P) is from 0.5 \% to 6.0 \%. The mean values of P and polarization position angle ($\theta_{P}$) with corresponding standard deviation are 2.7$\pm$1.0 \% and 179$\pm$11$^\circ$, respectively. I have chosen only those measurements for which the ratio of the P and the error in P ($\sigma_{P}$), P/$\sigma_{P}$, is $\geq$ 2 (this criterion provides the statistically significant number of data points). The observed polarization vectors are plotted on WISE 12 $\mu$m image of NGC 1893 shown in the left panel of the figure \ref{Fig:corrected}. The distribution of position angles with polarization fraction is shown in the right panel of the figure. Fig. \ref{Fig:P_PA_dist} shows the variation of amount of polarization and position with distances of the targets in left and right panels, respectively.

\begin{figure}
\centering
\resizebox{12cm}{10cm}{\includegraphics{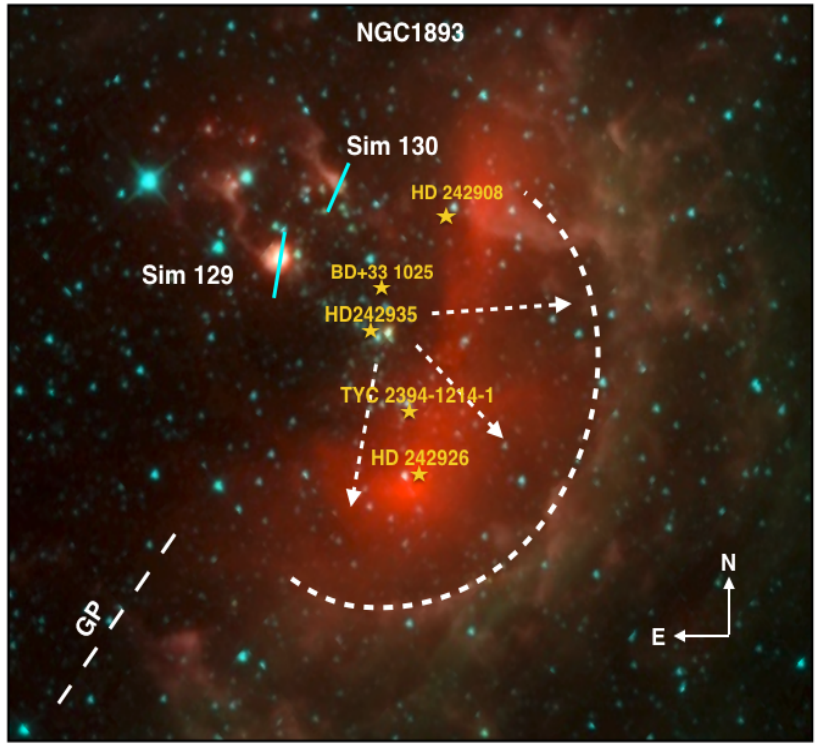}}
\caption{Figure shows the color composite (RGB) image of HII region IC\,410 region made using WISE 1 (3.4 $\mu$m), 2 (4.6 $\mu$m), and 4 (12 $\mu$m) bands data containing NGC\,1893 open cluster. The image is labeled with the position of nebulae Sim 129 and Sim 130. Locations of five O type stars including HD\,24293 are also shown in the image. The cyan line segments plotted on the nebulae represent the mean magnetic field orientation in these regions. The expanding region created by ionizing radiation from HD\,242935 is shown by dashed circle. The position of Galactic plane (GP) is shown using white dashed line and the north and east direction are also presented.}\label{Fig:labeled}
\end{figure}

Figure \ref{Fig:labeled} shows the color composite image (constructed using WISE bands 1, 2 and 4) of HII region IC\,410 containing NGC\,1893. The shape of dusty appearance of this HII region can be noticed in this figure. The expanding surrounding medium is clearly depicted in the image. The figure is labeled with positions of ionizing source HD\,242935 and other O type stars, and the Sim nebulae 129 and 130. The mean magnetic field directions in the nebulae are shown using the cyan color line segments. The image shows that the inherent magnetic fields in the region were mostly parallel to the Galactic plane. The shape of nebulae suggest that their surface material is mostly photo-evaporated by the ionizing radiation from central O type star leaving behind the smoky, thin shaped and elongated cloud structure \citep{1983ApL....23..119Z, 1991ApJS...77...59S}.

\subsection{Archival radio data}

\begin{figure}
\centering
\resizebox{10cm}{8cm}{\includegraphics{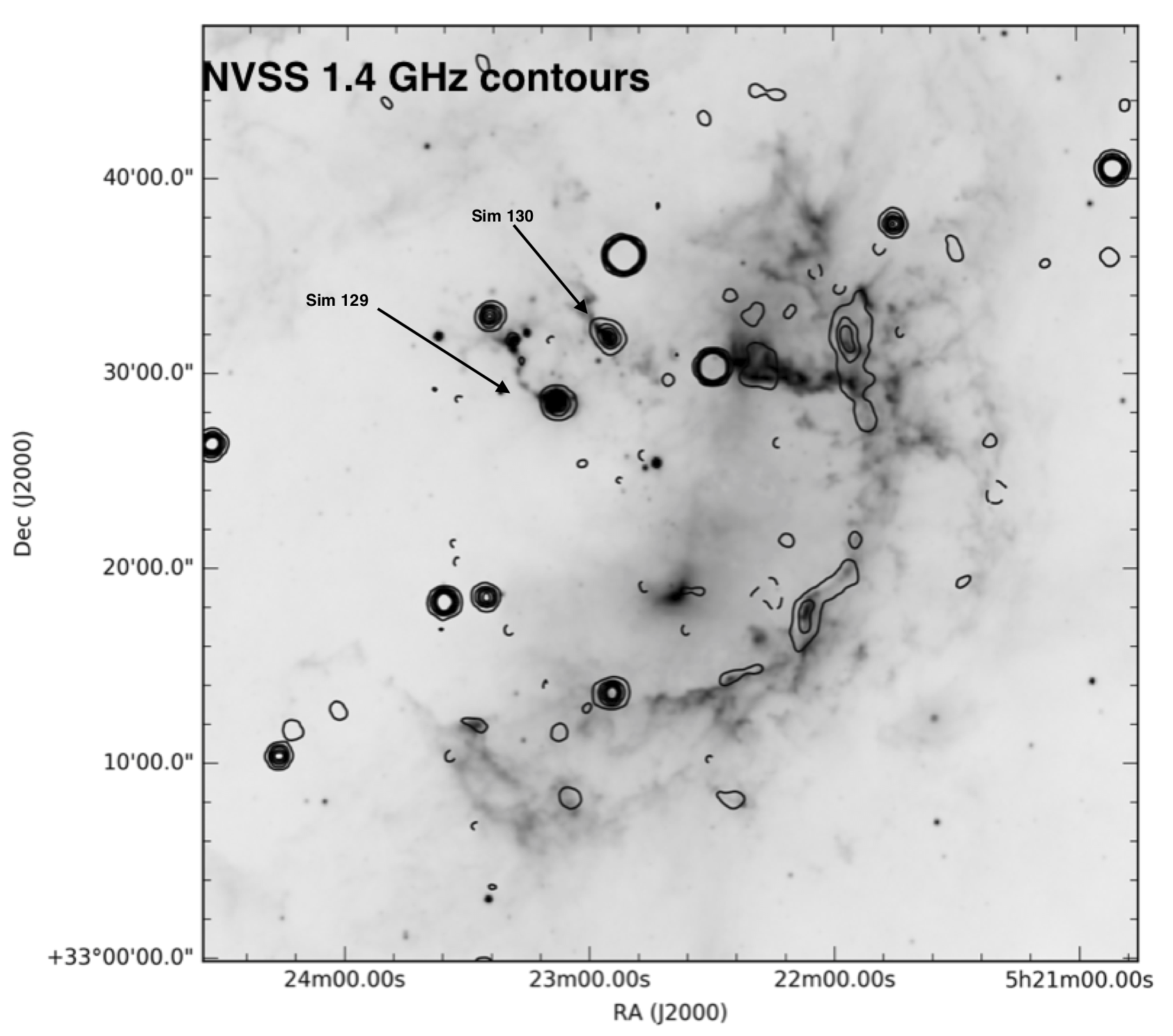}}
\caption{The NVSS 1.4 GHz contours shown on WISE 12 $\mu$m image of NGC 1893 containing Sim 129 and Sim 130 nebulae.}\label{Fig:radio610}
\end{figure}

The NVSS 1.4 GHz of the region in IC 410 are shown in figure \ref{Fig:radio610}. The radio 1.4 GHz is overplotted on WISE 12$\mu$m image. The NVSS map does not give much information due to its poor resolution and sensitivity but still the wall of HII region and location of two nebulae Sim 129 and 130 are clearly visible.  The resolution of this map is $45^{\prime\prime}$ at 1.4~GHz. The radio fluxes at nebulae Sim 129 and 130 are used to estimate some parameters discussed in section 4.3.

\subsection{Estimation of line width from CO molecular line observations}
\begin{figure}
\centering
\resizebox{15cm}{12cm}{\includegraphics{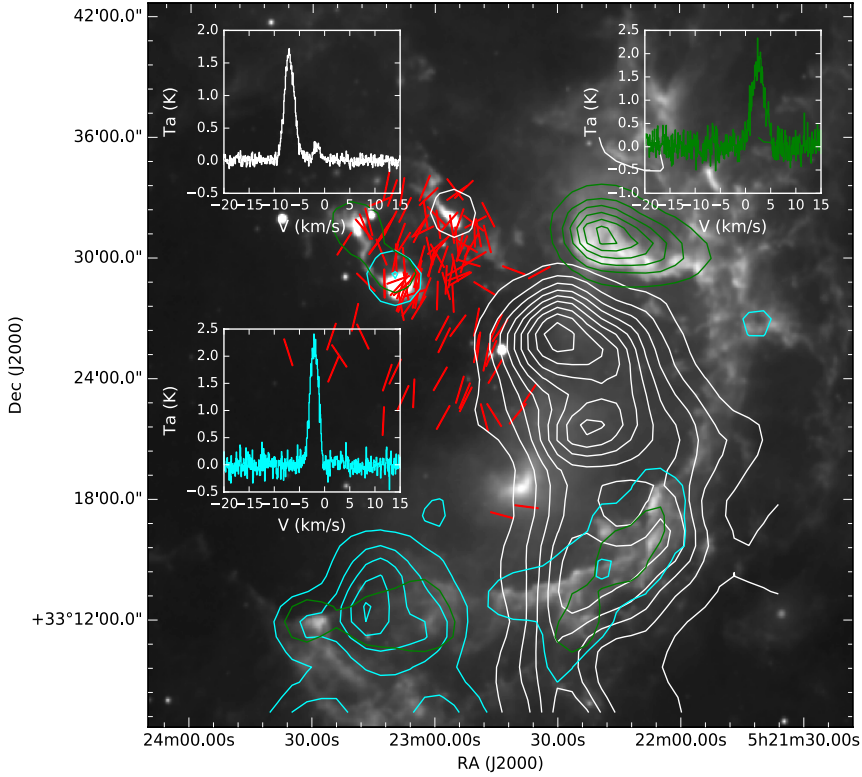}}
\caption{WISE 12$~\mu$m image of NGC\,1893 with overplotted CO (J=1-0) integrated intensity contours corresponding to three velocity components with $V_{LSR}$ -7.0$~\rm km~s^{-1}$ (white), -2.5$~\rm km~s^{-1}$ (cyan) and $V_{LSR}$ 2.5$~\rm km~s^{-1}$ (green) in the line of sight. The average spectra corresponding to these emissions are shown in the insets. The $x$ axes of the insets show the velocity and the $y$ axes shows the antenna temperature. The red segments are normalized polarization vectors (independent of polarization fraction) showing the magnetic field morphology in this region.}\label{Fig:ngc1893_CO}
\end{figure}

Figure \ref{Fig:ngc1893_CO} shows the CO contours overplotted on WISE 12$~\mu$m image containing IC410 region. I found three components of CO line with different $V_{LSR}$ $-$7.0$~\rm km~s^{-1}$, $-$2.5$~\rm km~s^{-1}$ and 2.5$\rm ~km~s^{-1}$ associated to this cloud which are shown using white, cyan and green color contours, respectively in figure \ref{Fig:ngc1893_CO}. The red line segments are normalized polarization vectors (where length of the vectors do not depend on polarization fraction) plotted to infer the magnetic field morphology in the region. I considered all the three components of CO emission since they coincide with locations of the cloud region where I carried out optical polarimetric observations. The insets in figure \ref{Fig:ngc1893_CO} show the average spectra of the CO emission. The width of spectral line is generally expressed as full width at half-maximum (FWHM) when the line profile is well represented by Gaussian-shape. From the OTF map in CO (J=1-0) molecular line towards NGC\,1893, I estimated the CO line width as it traces the low density region where optical polarization observations covers the cloud. The CO line width of the average spectra of the component with $V_{LSR}$ $-$7.0$~\rm km~s^{-1}$ is found to be 2.27$\pm$0.03$~\rm km~s^{-1}$ and that of the other two components with $V_{LSR}$ $-$2.5$~\rm km~s^{-1}$ and $V_{LSR}$ 2.5$~\rm km~s^{-1}$ are found to be 1.95$\pm$0.04$~\rm km~s^{-1}$ and 2.9$\pm$0.09$~\rm km~s^{-1}$, respectively. The average value of FWHM in these components is found to be 2.37$\pm$0.10$~\rm km~s^{-1}$. I measure CO line widths by fitting Gaussian profile to the line spectra.

\section{Discussion}

\subsection{Distance and interstellar polarization subtraction}

Since NGC\,1893 is located at a distance of 3.2 kpc \citep{2007MNRAS.380.1141S}, our polarimetric results towards NGC1893 may have some contamination introduced by ISM polarization in the line of sight. Therefore it is necessary to remove the foreground polarization contribution from our results. \citet{2011MNRAS.411.1418E} has presented multiband optical polarization observations of this region. They observed clusters located between 600 pc (NGC 2281) to 3.2 kpc (NGC\,1893) including  NGC\,2281, NGC\,1960 and Stock 8, along with NGC\,1664, to investigate the dust properties towards Galactic anticenter. They reported two dust layers at $\sim$170 pc and $\sim$360 pc distances based on distribution of $P_{V}$ and $\theta_{V}$ (polarization observations in V band). The first dust layer is found to have  $P_{V}$ $\sim$0.3-0.9 \% and $\theta_{V}$ $\sim$20-50$^\circ$ and the second layer is found to have $P_{V}$ $\sim1.0-1.9$ \% and $\theta_{V}$ $\sim$110-150$^\circ$. The maximum amount of polarization produced by the two layers is found to be $P_{V}$ $\sim$ 2.2 \%.

In order to remove the foreground interstellar polarization from our results, I used the polarization values of the cluster Stock 8 with Galactic coordinates of $\ell$ $=$ 173.37$^{\circ}$ and \textit{b} $=$ $-0.18^{\circ}$. The Galactic coordinates of NGC 1893 are $\ell$ $=$ 173.59$^{\circ}$ and \textit{b} $=$ $-1.68^{\circ}$. Since Stock 8 is at 2.5\,kpc \citep{2008MNRAS.384.1675J} distance, I assumed that it carries the maximum contribution of all the material present foreground to NGC 1893. I estimated the mean values of P and $\theta_{P}$ towards this cluster. I then estimated the Stokes parameters $q (=P\cos 2\theta$) and $u (=P\sin 2\theta$) using the polarization values towards Stock 8. I calculated the mean values of these Stokes parameters and named them as $Q_{fg}$ and $U_{fg}$. The mean values of foreground Stokes parameters $Q_{fg}$ and $U_{fg}$ are estimated to be 1.458 and -1.139, respectively. Then I calculated the Stokes parameters, $Q_{\star}$ and $U_{\star}$, of the target sources. The Stokes parameters $Q_{c}$ and $U_{c}$ representing the foreground corrected polarization of the target stars are calculated using the expression:
\begin{equation} \label{qu_star_ism}
Q_{c}=Q_{\star} - Q_{fg},\\
U_{c}=U_{\star} - U_{fg}
\end{equation} 

The corresponding foreground corrected degree of polarization $P_{c}$ and position angle $\theta_{c}$ of the target stars are calculated using the equations
\begin{equation} \label{ppa_star_ism}
P_{c}=\sqrt{(Q_{c})^2+(U_{c})^2},\\
\theta_{c}=0.5\times tan^{-1}\left(\frac{U_{c}}{Q_{c}}\right)
\end{equation}
\begin{figure}
\centering
\resizebox{12cm}{8cm}{\includegraphics{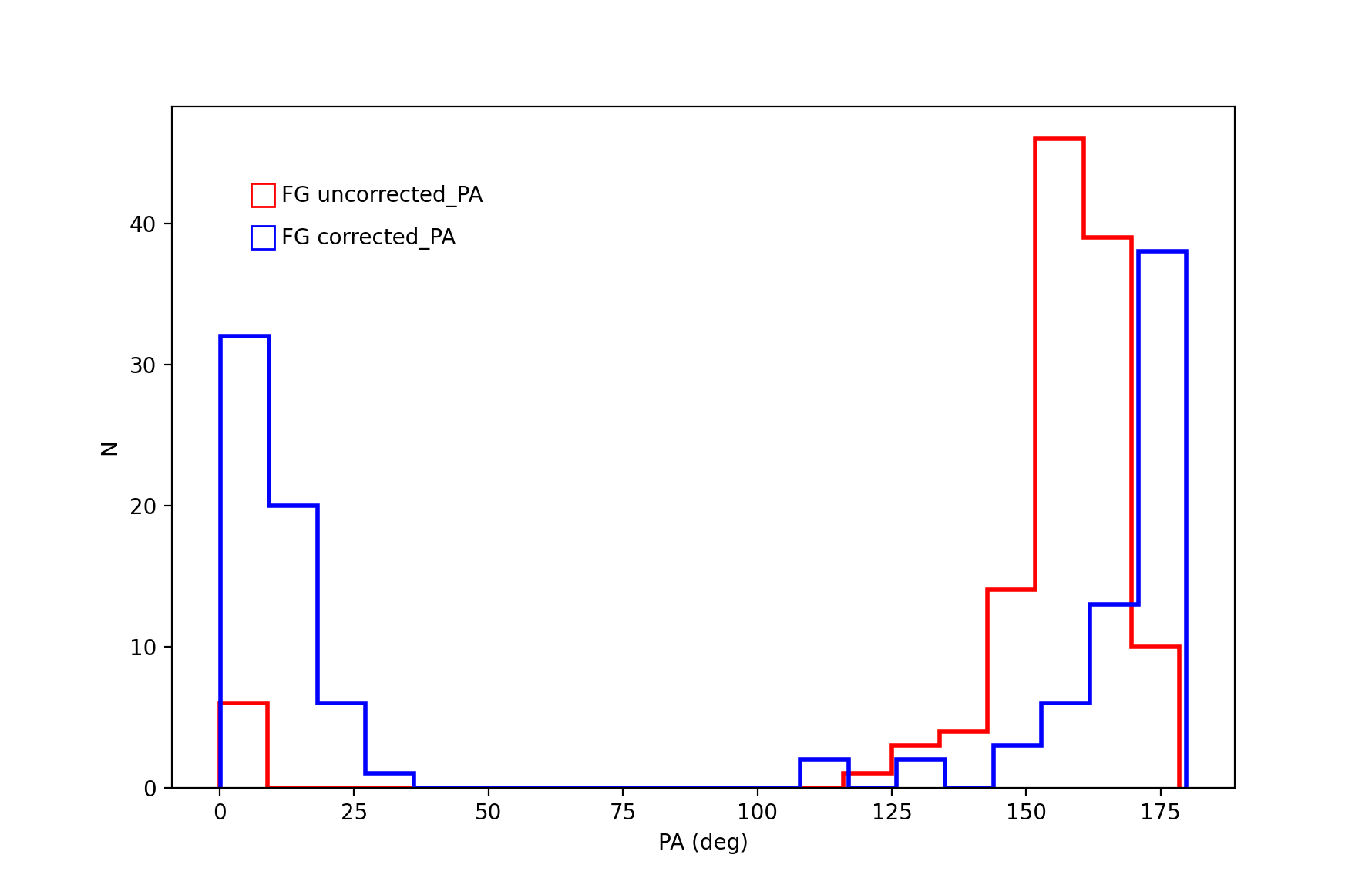}}
\caption{The histogram of foreground uncorrected and corrected position angles.}\label{Fig:hist_cor_uncor}
\end{figure}

The histograms of uncorrected and foreground corrected polarization position angles in figure \ref{Fig:hist_cor_uncor}. Both the populations have two components centered around $10^{\circ}$ and $150^{\circ}$ values. This suggests that these components are inherent in the cloud. The corrected polarization vectors are overlaid on WISE 12 $\mu$m image of NGC 1893 shown in the left panel of figure \ref{Fig:corrected}.

\subsection{Ricean bias correction in polarization}

\begin{figure}
\centering
\resizebox{13cm}{9cm}{\includegraphics{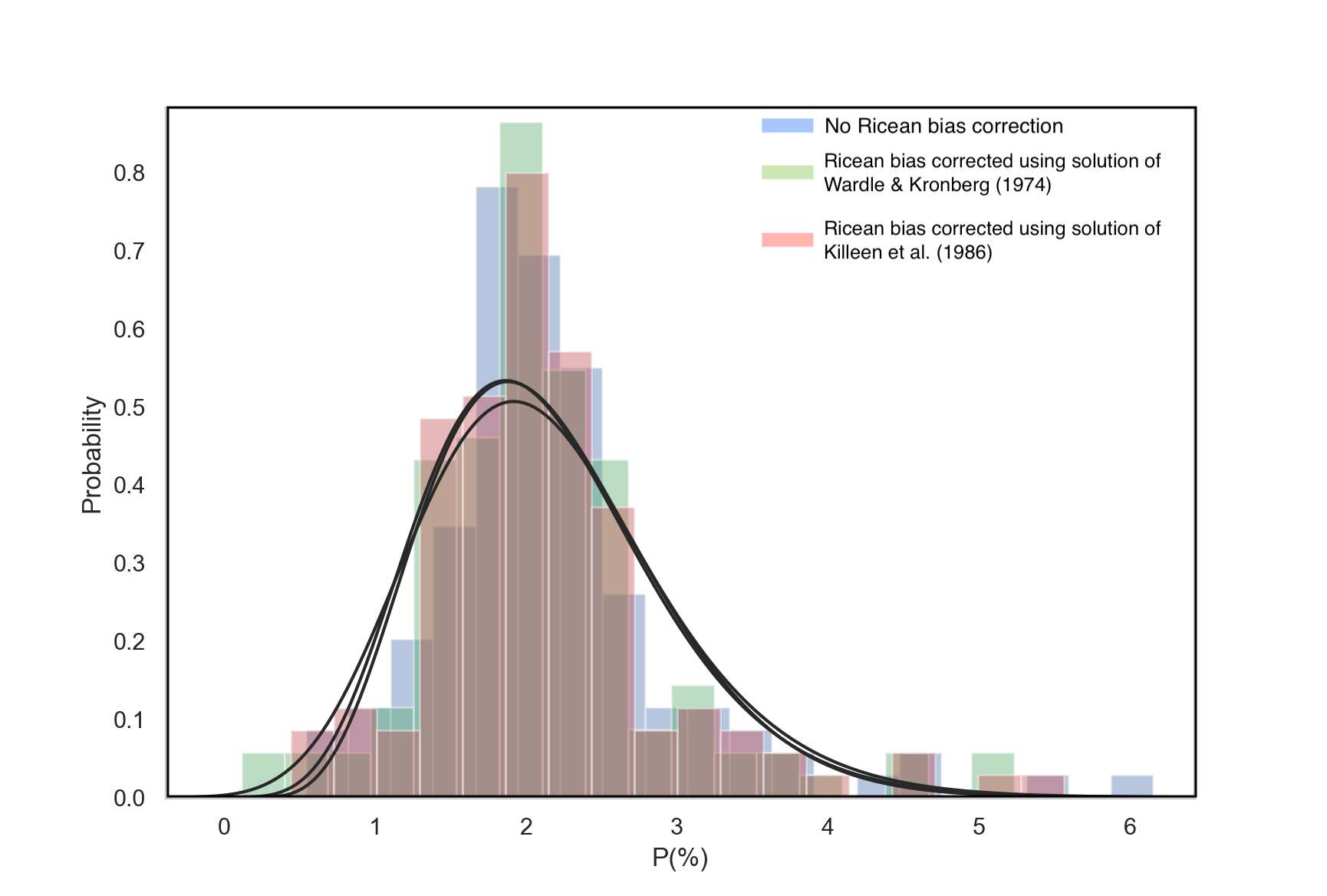}}
\caption{The histograms of polarization values without and with Ricean bias corrections \citep{2011JASS...28..267S}.}\label{Fig:ricean}
\end{figure}

After foreground contributions removal, I corrected the polarization values for Ricean bias using the approach mentioned in \citet{2011JASS...28..267S}. Ricean bias occurs in linear polarization measurements. \citet{2011JASS...28..267S} reviewed two methods for this correction. One method is developed by \citet{1974ApJ...194..249W} and the other similar method was developed by \citet{1986ApJ...302..306K}. In first method, a good approximate solution for Ricean bias correction is given by 
$I_{p}\sim$ $\sqrt{{I_{p_{obs}}}^{2} - {\sigma_{I_{p_{obs}}}^{2}}}$. Here, $I_{p}$, $I_{p_{obs}}$, and $\sigma_{p_{obs}}$ are bias corrected polarized intensity, observed value of polarized intensity, and associated uncertainties in $I_{p_{obs}}$, respectively. The other solution of ricean bias correction give by \citet{1986ApJ...302..306K} assumes above relation as $I_{p}\sim$ $I_{p_{obs}} - 0.5{\sigma_{p_{obs}}}^{2}/I_{p_{obs}}$. I used both these methods explained in \citet{2011JASS...28..267S} to perform the ricean correction in polarization values. The Gaussian fitted histograms of uncorrected, and corrected values using two methods is shown in Fig. \ref{Fig:ricean}. The difference in the corrected and un corrected values is not huge and both methods delivers almost similar values.

\subsection{Magnetic field geometry and strength}

\begin{figure*}
\centering
\resizebox{13cm}{12cm}{\includegraphics{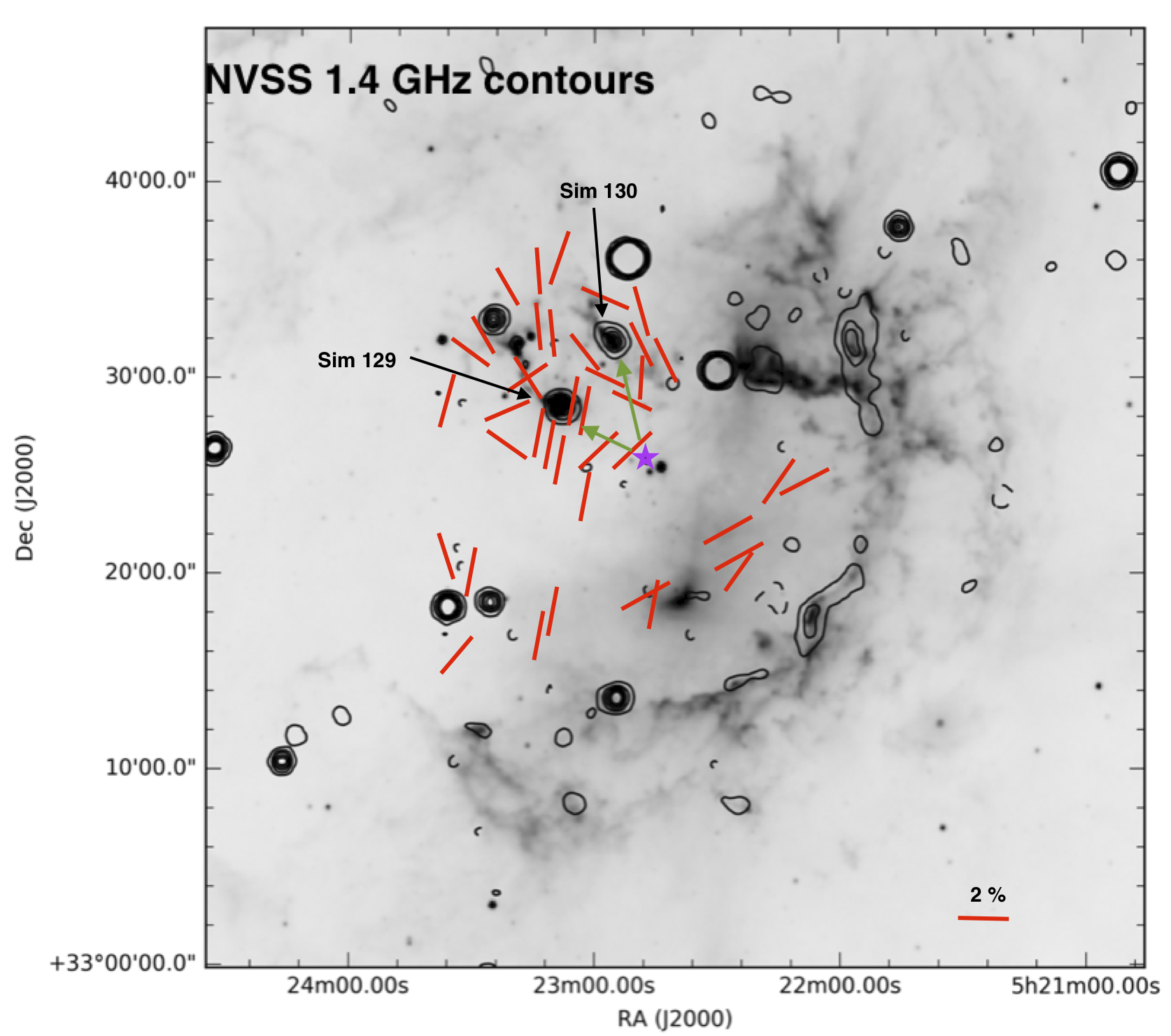}}
\caption{Same as figure \ref{Fig:radio610} with magnetic fields shown using normalized polarization vectors overplotted. The magenta star symbols shows the position of ionizing source HD 242935. The plane-of-sky directions of ionizing radiation toward nebulae is shown with green vectors.}\label{Fig:radio_pol}
\end{figure*}

\begin{figure*}
\centering
\resizebox{12cm}{10cm}{\includegraphics{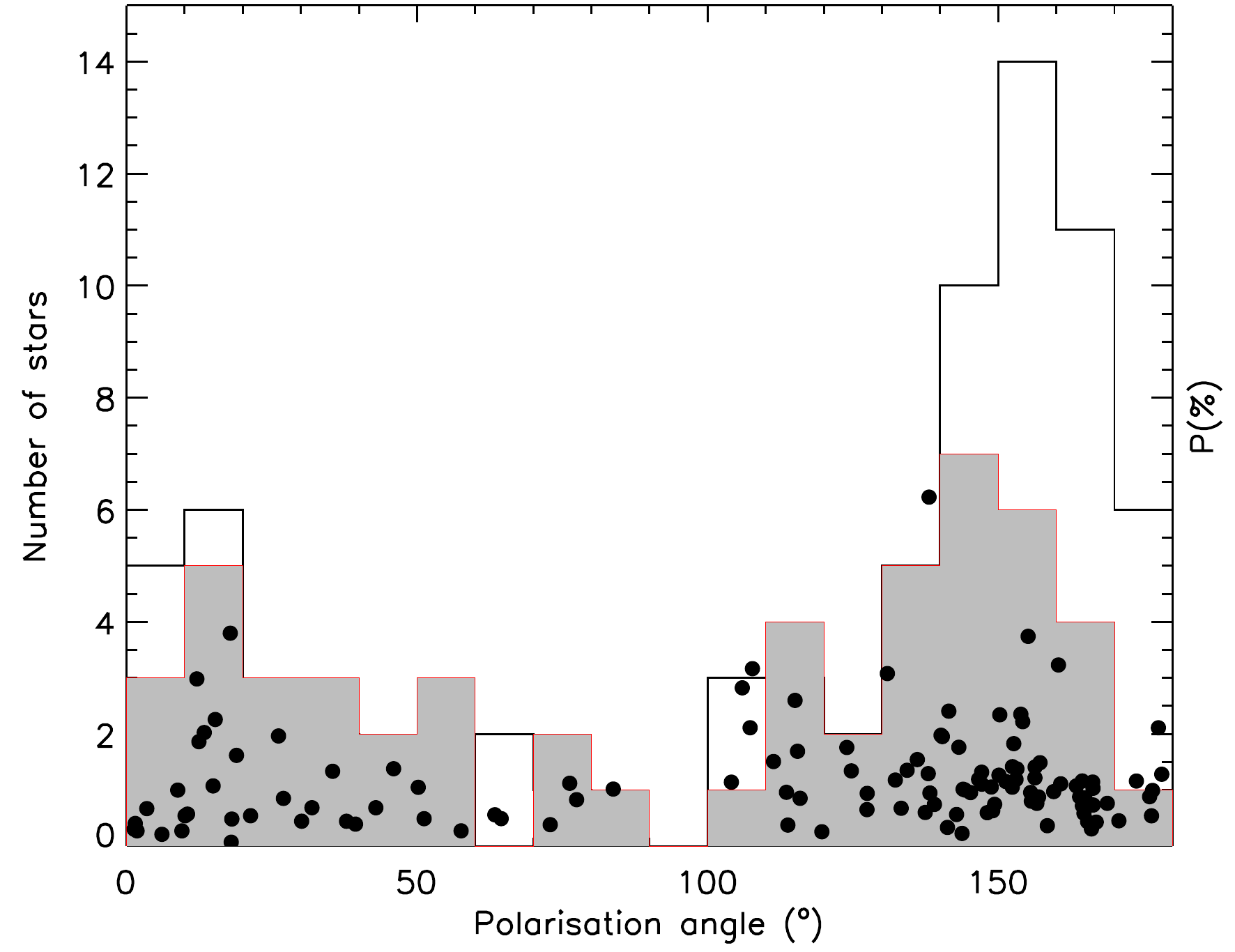}}
\caption{Histogram of the $\theta_{P}$ values is shown along with the distribution of P with $\theta_{P}$. The open histogram corresponds to the data in the surrounding of Sim 130 and the filled grey histogram represents the data in the vicinity of Sim 129.}\label{Fig:hist_corr_ngc1893}
\end{figure*}

The field geometry in NGC 1893 becomes complex after subtracting the foreground polarization component. Figure \ref{Fig:radio_pol} shows the normalized polarization vectors (i.e. length of the line segment is independent of the fraction of polarization) corresponding to the higher signal-to-noise-ratio (with P/$\rm \sigma_{P}$ $>$ 2.5) data overlaid with 1.4 GHz continuum contours on WISE 12 $\mu$m image of NGC\,1893 region.  Towards nebula Sim 129, it is apparent that the field lines are curved in the outward direction which is opposite to the morphology normally seen in some cometary globules (CGs) such as LBN 437 \citep{2013MNRAS.432.1502S}. Such field geometry has been seen in CG 30-31 complex by \citet{2011ApJ...743...54T}. This might be caused by the streaming off the gas from the head part of Sim 129 due to the presence of ionizing radiation from O type stars including HD 242935. This cloud is relatively closer to the ionizing radiation as compared to Sim 130 nebula. Therefore, the effect of the UV photons is relatively higher towards this nebula. The field lines may possibly be following the ionized gas photoevaporating in the outward direction assuming that the matter is coupled to the field lines. This will make the field lines curved towards the outward direction. A bimodal distribution of the position angles towards NGC 1893 can be noticed in figure \ref{Fig:hist_corr_ngc1893}. The direction of ionizing photons (shown with green arrows on Fig. \ref{Fig:radio_pol}. This direction of radiation is in the plane of the sky and obtained just by eye estimation) impinging on the head of Sim 129 is $\sim 60^\circ$ and the mean direction of magnetic fields in the cloud is $155^\circ$ (see figure \ref{Fig:radio_pol}). The offset of $95^\circ$ between the ionizing radiation and the magnetic fields suggest that the magnetic field lines are perpendicular to the incoming UV radiation. Such results can be compared to the three dimensional magnetohydrodynamics (MHD) simulations presented by \citet{2009MNRAS.398..157H}. These simulations suggest that the strong perpendicular magnetic fields resist any movement of the gas in directions perpendicular to the original field direction, whereas no such magnetic support exists to oppose gas motions along the field lines. As a result, the implosion of the globule by the ionization-driven shock front is highly anisotropic. In the case of weak magnetic fields perpendicular to the ionizing radiation direction, the flattening of the globule along the field lines is much lesser. In addition, no current sheet forms at the symmetry plane, and the photoevaporation flow is powerful enough to drive all ambient material. By comparing our observations towards Sim 129 nebula with the simulation, I found that the magnetic fields are relatively weaker and following the material in the streams of ionized gas from the head of nebula.

Whereas in Sim 130, the magnetic field geometry suggests that the fields lines are curved on the cloud head. It seems that the field lines are dragged by the ionizing radiation but in opposite direction than that of seen in Sim 129. This is similar to the field morphology seen in LBN 437 \citep{2013MNRAS.432.1502S}. The different field morphology in Sim 130 may be because its distance from ionizing source is relatively larger (projected distance of Sim 129 and Sim 130 from HD 242935 are 3.7 pc and 4.4 pc, respectively at a 3.2\,kpc distance of NGC1893) than that of Sim 129. The flux reaching the this cloud should be relatively lesser because of the condensation of ionized gas between the source HD 242935 and the nebula. Therefore the amount of ionization and ionizing gas streaming out of the nebula head should be lesser than that from Sim 129. The field morphology of Sim 129 may be more chaotic due to higher amount of ionizing photons and hence higher pressure compared to the magnetic pressure. This might not be the case in Sim 130. The histogram and distribution of position angle with degree of polarization is also shown in figure \ref{Fig:hist_corr_ngc1893}. There is a component of $\theta_P$ appearing in the figure \ref{Fig:hist_corr_ngc1893} peaking at around $\sim10^\circ$. The direction of ionizing radiation impinging on this nebula head is also $\sim10^\circ$. This suggest that the field lines become almost parallel to the ionizing radiation in Sim 130. MHD simulations towards such morphology of magnetic field are presented by \citet{2001MNRAS.327..788W} showing the curved field lines on the head of the cloud. The field lines may be dragged by the incoming ionizing radiation. There is a visible component of of $\theta_P$ appearing at $\sim150^\circ$ in both the nebulae but it is more prominent in Sim 129. This component is almost perpendicular to the average direction of ionizing radiation in plane-of-sky.

\begin{figure}
\centering
\resizebox{12cm}{6cm}{\includegraphics{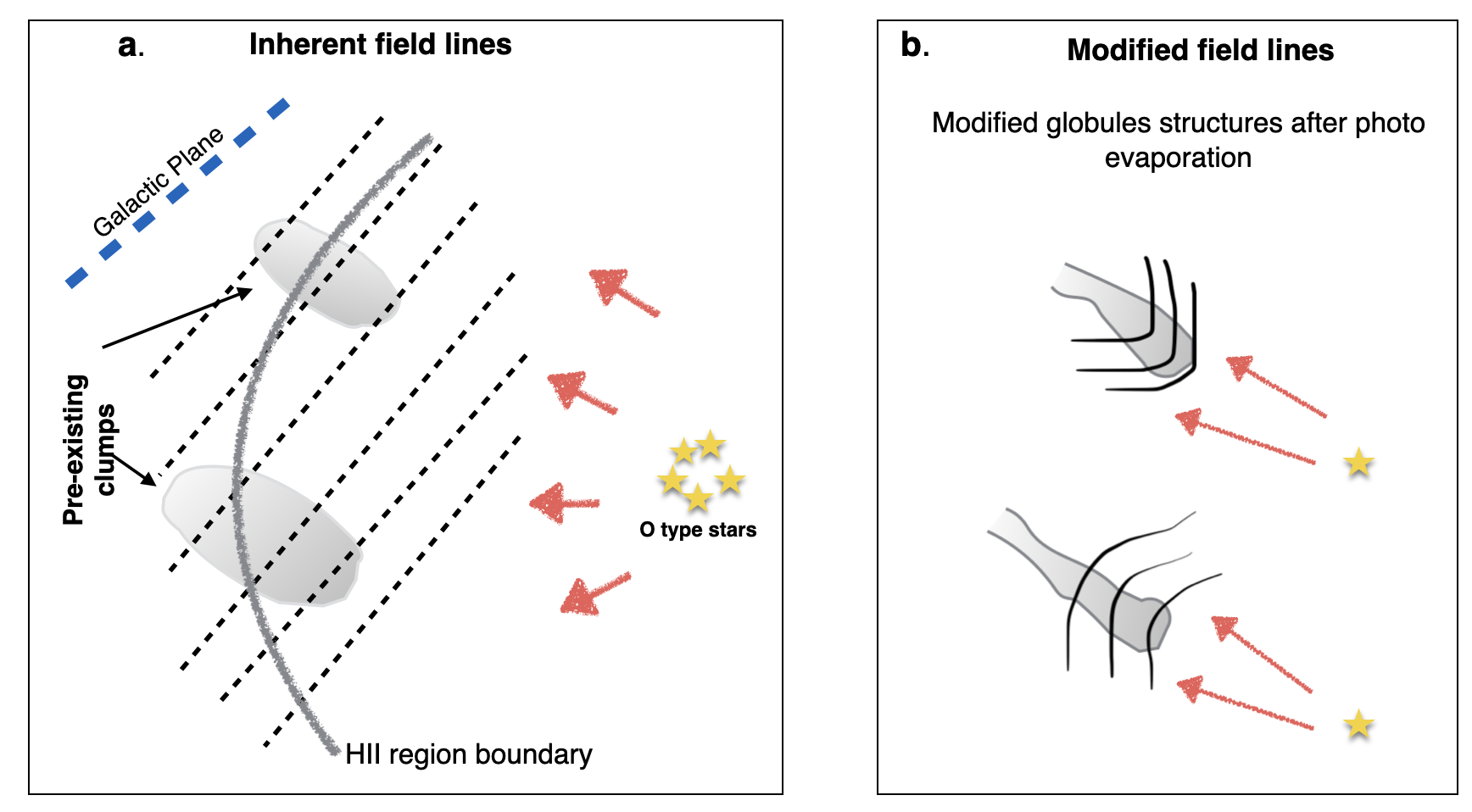}}
\caption{Figures a and b in left and right panels show the cartoon diagram depicting initial geometry of field lines with the location of the preexisting nebulae and the modification in the magnetic fields and nebulae structure after getting affected by the ionizing radiation from central sources, respectively.}\label{Fig:cartoon}
\end{figure}

I tried to investigate the structural evolution of nebulae Sim 129 and 130 in context of the magnetic fields by drawing a cartoon diagram shown in figure \ref{Fig:cartoon}. The left panel in this figure shows the initial geometry of magnetic fields in the region and the location of two possibly preexisting clouds on the periphery of HII region created by central O type stars including HD\,242935 (see figure \ref{Fig:labeled}). The ionizing radiation photo-evaporates the surface material of the two clouds causing a more elongated tadpole like  structures stretching radially away from the direction of radiation. The modified structures of the nebulae are shown in right panel of the figure \ref{Fig:cartoon}. The magnetic fields are modified on the tips of the nebulae based on their distances from the ionizing sources. The modified and curved fields lines are also shown by dashed curves in the right panel. 

I estimated the plane-of-the-sky ($\rm B_{POS}$) magnetic field strength in the region around Sim 129 and Sim 130 nebulae using updated Davis-Chandrasekhar-Fermi (DCF) relation ($B_{POS}=9.3\sqrt{n(H_{2})}\delta v/\delta\theta$; \citealt{1951PhRv...81..890D, 1953ApJ...118..113C}). Here $n(H_{2})$ represents the volume density of the clouds which is taken to $\sim 10^{3}\,cm^{-3}$, $\delta v$ is the velocity dispersion and $\delta\theta$ is the dispersion in $\theta_{P}$, corrected by the uncertainty in $\theta_{P}$\footnote{The uncertainty in the position angles is calculated by error propagation in the expression of polarization angle $\theta$, which gives, $\sigma_{\theta} = 0.5\times\sigma_{P}/P$ in radians, or $\sigma_{\theta} = 28.65\degree\times\sigma_{P}/P$ \citep[see;] []{Serkowski1974} in degrees.} (see details in \citealt{2001ApJ...561..864L, 2010ApJ...723..146F}). The average value of dispersion in position angles in both the components seen in Fig. \ref{Fig:hist_corr_ngc1893} is measured to be $\sim 13^{\circ}$. I measured the $\rm ^{12}CO$ (J=1-0) line width ($\Delta v$ = 2.3 $\rm km s^{-1}$) from our observations towards NGC\,1893 using TRAO. The corrected dispersion in position angle ($\delta\theta$) is estimated by following the procedure given by \citet{2001ApJ...561..864L} and \citet{2010ApJ...723..146F} using the standard deviation obtained from a Gaussian fit to the $\theta_{P}$. Using these values in DCF relation, I obtained a field strength of $\sim60~\mu$G around the two nebulae with a typical uncertainty ($\sigma B_{POS}$) of $\sim0.5B_{POS}$ \citep{2005mpge.conf..103C}.

\subsection{Pressure comparison in Sim nebulae}

When a cloud is exposed to high energy radiation from O and B stars, the surface of the cloud get ionized by the incident flux. This ionized gas streams away from the cloud and exert a pressure in the form of shock front on the cloud. Ionized boundary layer pressure can be calculated using the electron density estimated with the help pf radio flux.  This is important to understand the pressure balance in these clouds when we compare the IBL pressure to other pressure terms such as magnetic and internal molecular pressure. I estimated these pressures to investigate the competition between IBL and molecular pressure. This pressure balance investigation is important to understand because in pressure equilibrium condition, these clouds are likely to be shocked by photoionization. By using the radio intensities obtained towards the head parts of these nebulae (emission at the center of the head of nebulae in 1.4 GHz), I estimated the flux impinging on the nebulae head. I used the method adopted by \citet{2004A&A...426..535M} for estimation of the various parameters towards Sim 129 and Sim 130. I calculated the flux ($\phi$) reaching the Sim 129 and Sim 130 nebula using the general equations from \citet{1994A&A...289..559L}. They assumed a Gaussian intensity distribution of width $\theta$ (HPBW) and derive the relation (shown in eq. 6 of their paper) for a Rayleigh-Jeans brightness flux. \citet{2004A&A...426..535M} assumes an optically thin emission and the region is in photoionization equilibrium. Adopting the similar approach, I used the following relation to estimate the impinging flux.

\begin{equation}
\phi = 1.24\times10^{10} S_{\nu} {T_{e}}^{0.35} \nu^{0.1} \theta^{-2}
\end{equation}

Where the symbols $S_{\nu}$, $T_{e}$, $\nu$ and $\theta$ represent the integrated radio flux in Jy/beam, effective electron temperature of the ionized gas in K, frequency of the emission in GHz and angular diameter in arcsec over which the emission is integrated. The average value of $S_{\nu}$ over Sim 129 and 130 are found to be 0.021 and 0.024 Jy/beam, respectively. I used typical value of electron temperature $T_{e}$ as $10^{4}$\,K. Considering the flux obtained in 1.4~GHz emission from NVSS and an angular diameter of $\sim$ 50 arcsec, I calculated the flux ($\phi$) reaching the nebulae. The estimated values of $\phi$ towards Sim 129 and Sim 130 are found to be $\rm 6.2\times 10^{9} cm^{-2} s^{-1}$ and $\rm 5.1\times 10^{9} cm^{-2} s^{-1}$, respectively. The subsequent values of electron density estimated towards Sim 129 and Sim 130 are also calculated as :

\begin{equation}
n_{e} = 122.41 \times \sqrt{\frac{S_{\nu} {T_{e}}^{0.35} \nu^{0.1} \theta^{-2}}{\eta R}}
\end{equation}
 
 Where the symbols are same as shown in equation above and $\eta$ is the effective thickness of the ionized boundary layer (IBL) as a fraction of the cloud radius (IBL; typically found as $\sim 0.2$ from \citealt{1989ApJ...346..735B}). R is the radius of the cloud in parsec. The values of electron densities are estimated to be $\sim$250 $cm^{-3}$ and $\sim$230 $cm^{-3}$ towards Sim 129 and Sim 130, respectively. The estimated values of IBL pressure ($\rm P_{i}/k_{B}$; where $\rm P_{i}$ is the ionizing gas pressure and $\rm k_{B}$ is the Boltzmann constant) are found to be $\rm 25\times 10^{5}cm^{-3} K$ and $\rm 23\times 10^{5}cm^{-3} K$ in Sim 129 and Sim 130, respectively. These results are comparable to the values found in some bright-rimmed clouds (BRCs) studied by \citet{2004A&A...415..627T}. To examine the pressure balance between in nebulae, I compared the IBL pressure with internal pressure. For estimating the internal pressure, I used our $\rm C^{18}O$ (J=1-0) line observations towards these nebulae using TRAO. I used the turbulent velocity dispersion ($\rm \sigma$) of $\rm C^{18}O$ (J=1-0) line and molecular gas density ($\rm \rho_{m}$). I used the relation ($\rm P_{m} \simeq \sigma^{2}\rho_{m}$) adopted by \citet{2004A&A...426..535M} for calculating the molecular pressure in BRCs.  The internal pressure values from molecular line data towards Sim 129 and 130 are found to be $\rm 23\times10^{5}cm^{-3}$K and $\rm 21\times 10^{5}cm^{-3}$K, respectively. The IBL pressure is found to be comparable to the internal pressure in the nebulae which suggests that these nebulae are in a state of pressure balance.
 
 \subsection{Triggered star formation in Sim nebulae} 
 
 Recently \citet{2018MNRAS.477.1993L} have found the evidences of feedback-driven star formation in NGC 1893 using high-resolution spectroscopy of stars and gas in the young open cluster NGC 1893. Their results suggest that newborn stars and the tadpole nebula Sim 130 are moving away from the central cluster containing O-type stars. They estimated a sequential star formation time scale of $\sim$1 Myr within a 9 pc distance. They found that $\sim$ 18 percent of the total population of the stars are formed by feedback from massive stars. This finding suggest that this process might have helped in the formation of OB association. The results presented by \citet{2018MNRAS.477.1993L} also support the self-regulating star formation model \citep{1977ApJ...214..725E}.

\subsection{Kinematics of Sim nebulae} 
Recently, \citet{2020A&A...633A..27O} published the kinematics of these two nebulae Sim 129 and 130 using multiple gas tracers such as $\rm ^{12}CO (J=3-2)$, $\rm HCO^{+}$, $\rm C_{2}H$, HNC, and HCN J=4-3 transitions. They detected kinematic signatures of infalling gas in the $\rm ^{12}CO (J=3-2)$ and $\rm C_{2}H$ J=4-3 spectra toward Sim 129. Whereas, they noted that the possible star formation activity in Sim 130 has not started yet when they analyzed the HCN/HNC integrated ratio of approximately three. Our profile of $\rm ^{12}CO (J=1-0)$ gas from TRAO observations in this region is mostly matching with their higher profile from high transition CO gas. This suggest that the gas motion is almost intact and coherent in low and high density regimes. Ortega et al. also found that Sim 129 may be in much evolved stage with star formation activity inside the head part of the globule. This finding was supported by their non-detection of optically thin tracers such as HCN and HNC in Sim 129. This could be due to the radiation from the newly formed star(s) in Sim 129. 

\section{Summary}\label{conclusion}

The nebulous emission and the structures in NGC\,1893 region are revealed using 1.4~GHz observations from NVSS archival data. Sim 129 and Sim 130 are the nebulae undergone photoionization in the vicinity of five O type stars including HD\,242935. The magnetic field lines towards the head of the nebulae seem to be intact towards both nebulae resulting into a curved field orientation. This happens generally when the field lines are perpendicular to the ionizing photons. Some polarization vectors are found to follow the radio emission suggesting magnetic fields are dragged away by the ionizing radiation from the central O type star. The $\rm ^{12}CO (J=1-0)$ molecular line observation are used to measure the velocity dispersion required for estimating the magnetic field strength using DCF relation. The plane of the sky magnetic field strength is found to be $\sim$ 60$\mu$G with uncertainty of 0.5 times the estimated strength towards the region covered with polarization observations. The IBL pressure is found to be comparable to the internal pressure in Sim nebulae suggesting that the nebulae are in a state of photoionization induced pressure balance. Recently the evidences of feedback triggered star formation are found in NGC\,1893 supporting the self-regulating star formation model.

\normalem
\begin{acknowledgements}
I thank anonymous referee for a very constructive report. I also thank NSF funding 1715876 for a partial supporting this research. I also acknowledge the KASI postdoctoral funding while this project was in progress in 2017. I thank Dr. Chang Won Lee and Dr. Maheswar for discussion on this project. 
\end{acknowledgements}
  
\bibliographystyle{raa}
\bibliography{ms}

\end{document}